\documentstyle[aps,twocolumn,prb,epsf]{revtex} 
\begin{document}
\draft
\twocolumn[    
\hsize\textwidth\columnwidth\hsize\csname @twocolumnfalse\endcsname    

\title{ Kondo effect in Ce$_{x}$La$_{1-x}$Cu$_{2.05}$Si$_{2}$ intermetallics}
\author{I. Aviani, M. Miljak, and V. Zlati\'c}
\address{Institute of Physics,\\
Bijeni\v{c}ka cesta 46, P.O. Box 304,\\
HR-10 001 Zagreb, Croatia}
\author{K.-D. Schotte} 
\address{Institut f\"ur Theoretische Physik, Freie Universit\"at Berlin, 
Arnimallee 14, Berlin D-14195 Germany}
\author{C. Geibel and F. Steglich}
\address{Max-Planck Institute for Chemical Physics of Solids, D-01187 Dresden, Germany}
\date{\today}
\maketitle

\widetext
\begin{abstract}
  The magnetic susceptibility and susceptibility 
  anisotropy of the quasi-binary alloy system 
  Ce$_{x}$La$_{1-x}$Cu$_{2.05}$Si$_{2}$ have been studied 
  for low concentration of Ce ions. 
  The single-ion description is found to be valid for $x\leq 0.09$.
  The experimental results are discussed in terms of the degenerate 
  Coqblin-Schrieffer model with a crystalline electric 
  field splitting $\Delta\simeq 330$ K. 
  The properties of the model, obtained by combining the lowest-order scaling 
  and the perturbation theory, provide a satisfactory 
  description of the experimental data down to 30 K. The experimental results 
  between 20 K and 2 K are explained by the exact solution of the 
  Kondo model for an effective doublet. 
\end{abstract}


]      

\narrowtext

\subsection{Introduction\label{introduction}}
The intermetallic compound Ce$_{x}$La$_{1-x}$Cu$_{2.05}$Si$_{2}$ 
has been studied for more than two decades,  but 
its properties are still not completely understood. 
For a large concentration of Ce ions the compound shows heavy 
fermion superconductivity \cite{steglich.79} and small-moment
antiferromagnetism. \cite{Bruls94,steglich97}  
At intermediate concentrations, non-Fermi-liquid features are
found, \cite{Andraka94} while for small $x$ one observes anomalies
due to the crystalline electric field splitting (CF) and the Kondo
effect. \cite{onuki.87b,aliev.84,ocko.99} 
The change of the properties of the compound induced by Ce
doping is an important issue that remains to be clarified, both
experimentally and theoretically.

Here we present magnetic susceptibility data of 
Ce$_{x}$La$_{1-x}$Cu$_{2.05}$Si$_{2}$ intermetallics 
for small concentrations of Ce ions, $x\leq 0.09$, 
and show that the magnetic moment of Ce  depends on 
temperature and is very anisotropic. \cite{aviani.99}   
The samples we consider are sufficiently dilute for the 
interaction between the Ce ions to be neglected. We find 
the susceptibility of a single Ce ion embedded in a tetragonal host 
by studying the effects of Ce concentration. 
The magnetic anomalies are accompanied by transport anomalies, as 
indicated by large peaks in the thermoelectric power and a 
logarithmic increase of the electric resistance. \cite{ocko.99} 
This behavior is typical of exchange scattering on CF-split ions
and we base the theoretical analysis on the degenerate Coqblin-Schrieffer 
model \cite{coqblin.69} in which the two lowest CF states 
are coupled  by exchange interaction to conduction states.

The CF parameters of dilute Ce alloys are obtained by analyzing the 
high-temperature magnetization data with the usual CF theory which 
neglects the coupling between the $f$  electrons and the conduction states,   
and considers just an isolated Ce ion in a tetragonal environment 
(for details see the Appendix). 
The magnetic anisotropy is explained by taking  $\Delta \simeq 330$ K 
for the CF splitting between the doublet ground state and a pseudo 
quartet, formed by the two closely-spaced excited doublets, 
and  taking $\eta=0.816$ for the relative weight of the CF orbitals. 
These values are consistent with the high-temperature neutron scattering 
data on CeCu$_{2}$Si$_{2}$ single crystals.
\cite{holland-moritz.96,goremychkin.93}  
However, the CF theory does not describe correctly the response of 
the high-temperature local moment, and it also fails to explain 
the rapid reduction of the local moment at intermediate temperatures 
and the loss of anisotropy at low temperatures. 
These characteristic features of the magnetic response of Ce ions in 
metallic hosts can only be obtained by coupling the $f$  states 
to the conduction band. 

In this paper we show that the magnetic properties of dilute 
Ce$_{x}$La$_{1-x}$Cu$_{2.05}$Si$_{2}$ compounds can be explained by 
describing the interaction between the Ce ions and the metallic electrons 
by the Coqblin-Schrieffer model with the CF splitting. 
The high-temperature susceptibility obtained by the lowest-order 
perturbation theory agrees very well with the experimental data, 
provided we renormalize the exchange interaction by the `poor man's 
scaling'.\cite{anderson.70,yamada.84,hanzawa.85}
For large CF splitting, the scaling theory generates two relevant 
low-energy scales, $T_K\ll T_K^H$, instead of one without splitting. 
The behavior of the anisotropic susceptibility above 30 K 
is accounted for by choosing $T_K=8.5$ K and $T_K^H=100$ K. 
The two low-energy scales, which differ by an order of magnitude, 
explain the reduction of the local moment from a high-temperature 
sextet to a low-temperature doublet, and lead to a qualitative 
explanation of the transport data.\cite{ocko.99,zlatic.00}
The low temperature properties of the Coqblin-Schrieffer model 
can not be calculated by scaling. However, for $\Delta=330$ K 
and $T_K^H=100$ K, the occupation of the excited CF states 
below 20 K is negligibly small. Thus, we approximate the $f$ state 
by an effective doublet, and describe the low-temperature properties 
by an effective spin-1/2 Kondo model with Kondo scale $T_K$. 
The exact renormalization group results for the susceptibility\cite{wilson.75} 
agree very well with the experimental data. 

The paper is organized as follows. First we describe the 
sample preparation and provide the details of the measurements.
Then we analyze the susceptibility data obtained by the 
Faraday magnetometer on powdered samples and the anisotropy data 
obtained by the torque magnetometer on polycrystalline samples,
and show that the observed anomalies can not be explained 
by the CF effects due to an isolated $f$ state. 
The coupling between the $f$ state and the conduction band, as described
by the Coqblin-Schrieffer model with a CF splitting, is introduced next.  
Then, we discuss the susceptibility and the magnetic anisotropy 
of an excited quartet separated from the ground state doublet by 
the energy $\Delta$. 
Finally, compare the theoretical results the experimental data. 
The CF calculations for Ce ions in a tetragonal environment are presented in the Appendix.  
\subsection{Experimental details and data analysis\label{experimental}}
\subsubsection{The samples             \label{experimental_samples}}

The polycrystalline  Ce$_{x}$La$_{1-x}$Cu$_{2.05}$Si$_{2}$ samples 
with low Ce-content $(x \leq  0.09)$ were prepared using a two-step 
procedure in order to enhance composition control and improve 
homogeneity. In the first step, appropriate
amounts of pure elements were arc-melted to get master ingots with
compositions $x = 0$ and $x = 0.1$. In the second step, part of these master
ingots were melted together in an appropriate ratio to obtain samples with
$0.01 \leq x \leq 0.09$. Samples with larger Ce-content were prepared in a single
step directly from pure elements. All samples were annealed for 40 hrs
at 700 $^\circ$C, then for 80 hrs at 950 $^\circ$C, followed 
by a slow cooling to 700 $^\circ$C  within 72 hrs. 
The X-ray powder diffraction patterns showed only reflections
of the ThCr$_{2}$Si$_{2}$ structure. Both lattice parameters $a$ and $c$, as well 
as the unit cell volume, decrease linearly with increasing Ce content,  
the decrease being quite pronounced for $a$  but rather weak 
for $c$. \cite{ocko.99}
This preparation procedure combines all the experience gained in our
laboratory within the last twenty years upon the investigation of 
CeCu$_2$Si$_2$ 
and related alloys. More detailed information about the preparation 
process can be obtained directly from the authors.\cite{geibel}

The temperature dependence of the susceptibility $\chi(x,T)$ is 
found by measuring the fixed powders weighing about 20 mg. 
The data analysis shows 
that $\chi(x,T)$ is dominated by the contribution due to 
Ce ions, but that the samples with the same nominal Ce 
Concentration \cite{systematic_error_1} show slightly 
different susceptibilities at low temperatures. 
This is demonstrated in Fig.\ 1, where $\chi(x,T)$ is plotted 
as a function of temperature for two pairs of typical samples. 
The susceptibility difference for the two samples with the 
same nominal concentration $x=0.07$ grows from about 
$2\times 10^{-6}$$(emu/mol)$ at 300 K to about 
$200\times 10^{-6}$$(emu/mol)$ at 2 K, while $\chi(x,T)$ 
increases in the same temperature interval only about 10 times. 
Similar behavior is seen in other samples, and we take that 
as an evidence that the susceptibility difference shown in 
Fig.\ 1 is not due to an inhomogeneous distribution of Ce ions. 
The slight variation of the functional form of $\chi(x,T)$ is most 
pronounced at low temperatures (see also Fig.\ref{fig_sus_ion}), 
and that can be explained in terms of a small variation in the 
Kondo temperature $T_K$. 
The inset in Fig.\ 1 shows the data for the two samples with 
nominally 7 at\% of Ce, plotted versus $T/T_K$. The low--temperature 
data follow the same curve, provided we use $T_K=7.6$ K and  
$T_K=9.3$ K for the data represented by the open and filled triangles, 
respectively.\cite{systematic_error_2} 
However, at intermediate temperatures, the data represented by the open  
and filled symbols can not be represented by a single function of $T/T_K$.
The Kondo temperature of various samples might differ because 
of small deviations of the actual Cu concentration from the nominal one
and the associated chemical pressure effects.  
\begin{figure}[tbp]
\epsfxsize=3.0in \epsffile{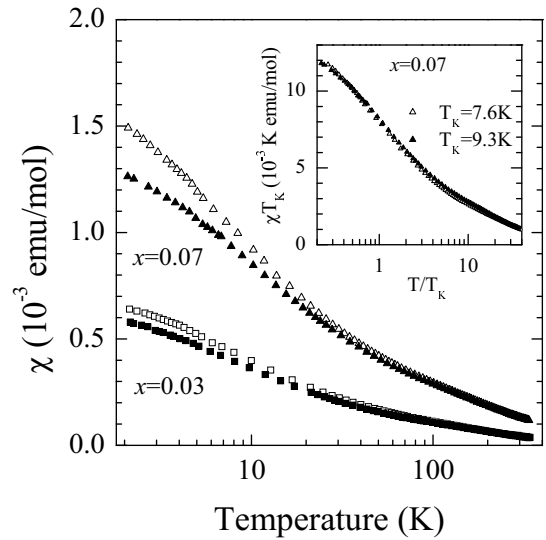}
                                                       \label{fig_error}
\caption{$\chi(x,T)$ plotted versus 
temperature for two pairs of Ce$_{x}$La$_{1-x}$Cu$_{2.05}$Si$_{2}$ samples. 
The triangles correspond to $x=0.07$ and 
the squares to $x=0.03$, respectively.  
The inset shows $\chi(x,T)\cdot T_K$ for the two $x=0.07$ samples, 
plotted versus $T/T_K$, with two different values of $T_K$.}
\end{figure}
The samples used for the torque measurements are small polycrystallites 
of irregular shape.
The uniaxial symmetry of the unit cell allows one to obtain the 
intrinsic anisotropy even in the absence of perfect alignment 
of the crystallites (for details see below). 
\subsubsection{The susceptibility\label{experimental_susceptibility}}
The absolute value of the average magnetic susceptibility $\chi (x,T)$ is
measured for Ce$_{x}$La$_{1-x}$Cu$_{2.05}$Si$_{2}$ samples on a
high-precision Faraday balance for temperatures between 2 K and 350 K, for
the magnetic field up to 9 kOe, and for Ce concentration ranging from 1 to 9
at\%. The magnetization of all samples, measured at 300 K, 77 K, and 2 K, is
found to be a linear function of the applied field. Susceptibility is
defined as $\chi (x,T)=M/H$ $[emu/mol]$. The experimental results are
plotted in Fig.\ 2 as a function of temperature and in Fig.\ 3 as a function
of concentration. 
Fig.\ 3 shows that the susceptibility increase linearly with Ce concentration,  
and we take that as an evidence that the Ce-Ce interaction is small.

\begin{figure}[tbp]
\epsfxsize=3.0in \epsffile{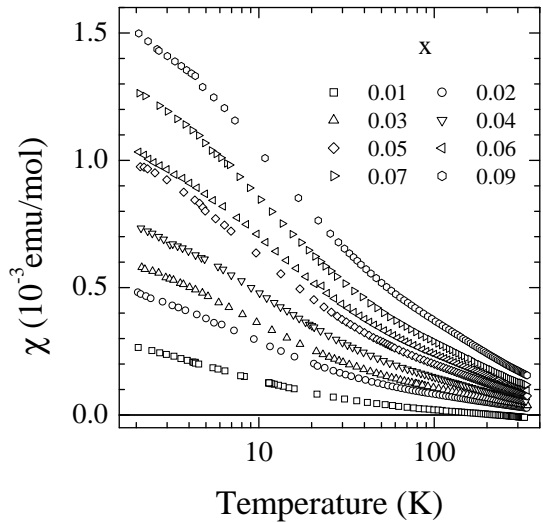}
                                                \label{fig_sus_t}
\caption{Susceptibility of Ce$_{x}$La$_{1-x}$Cu$_{2.05}$Si$_{2}$ 
plotted versus temperature on a logarithmic scale for $x\leq 0.09$.}
\end{figure}

The data analysis shows that the high--temperature susceptibility of 
the samples with the lowest concentration of Ce is diamagnetic, 
i.e., the response of very dilute alloys is dominated by a background. 
Thus, the effects of the alloying on the matrix 
in which the Ce ions fluctuate should not be a priory neglected. 
We also notice that the low--temperature 
susceptibility has a Curie-like upturn due to some unspecified magnetic 
impurities that are immanent to the rare earths, 
and for very dilute alloys this upturn becomes relatively large.
Because the concentration of these unspecified magnetic impurities differs 
in each sample, and because the alloying changes the background, 
we do not define the single--ion Ce--response as a difference between the 
susceptibility of Ce$_{x}$La$_{1-x}$Cu$_{2.05}$Si$_{2}$ and LaCu$_{2.05}$Si$_{2}$.  
To minimize the systematic error due to sample preparation  
we define the susceptibility of a Ce--free compound as a statistical average 
obtained from Fig.\ 3, and find the single--ion contribution by the 
following procedure.

\begin{figure}[tbp]
\epsfxsize=3.0in \epsffile{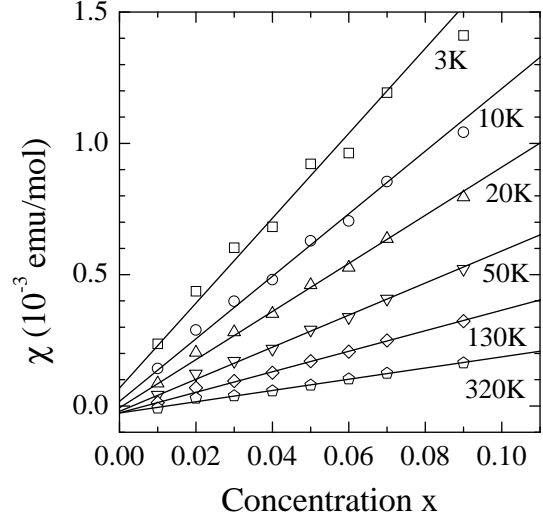}
                                                   \label{fig_sus_x}
\caption{Susceptibility of Ce$_{x}$La$_{1-x}$Cu$_{2.05}$Si$_{2}$ plotted
versus concentration for various temperatures.}
\end{figure}

We assume that the total susceptibility of a given sample, 
which is shown in Figs.\ 2 and \ 3,  comprises two contributions: 
a single-ion Ce susceptibility, $\chi _{ion}$, and a
concentration-dependent background, 
$
\chi _{B}(x,T)
\simeq 
\chi (0,T)
+x\cdot 
\left[ 
\partial \chi _{B}(x)/\partial x
\right]_{x=0}.  
$
We consider only the lowest order concentration effects and 
express the measured susceptibility of a given sample as, 
\begin{equation}
                           \label{sus_x&T}
\chi (x,T)
=\chi(0,T)
+x\left\{ 
     \chi_{ion}(T)+
       \left[ 
          \partial \chi_{B}(x)/\partial x
       \right]_{x=0}
  \right\}.  
\end{equation}
To find $\chi _{B}(x,T)$ we notice that the susceptibility data shown in
Fig.\ 3 are linear in concentration for all temperatures,
and use the $x=0$ intercept of the straight lines to define the
statistically averaged background susceptibility of the Ce-free 
matrix, $\chi (0,T)$.
The values obtained in such a way are shown in Fig.\ 4, 
together with the least-squares fit based on the expression 
\begin{equation}
                 \label{x=0_background}
\chi(0,T)
=
\chi _{0}+\frac{C_{imp}}{T-\Theta_{imp}}, 
\end{equation}
where $\chi _{0}$ describes the constant diamagnetic contribution 
of the stoichiometric compound LaCu$_{2.05}$Si$_{2}$
and the Curie-Weiss term describes magnetic contamination. 
The experiment gives, 
$\chi _{0}=-3.1\cdot10^{-5}$ $(emu/mol)$, 
$C_{imp}=5.9\cdot 10^{-4}\;$K ($emu/mol$), and 
$\Theta_{imp}=-2.8$ K. 
To estimate the level of contamination we assume 
$C_{imp}=(N_{imp}/3k_{B})g^{2}\mu _{B}^{2}j(j+1)$, where $\mu _{B}$ is Bohr
magneton, $g$ is the gyromagnetic factor and $k_{B}$ is Boltzmann constant.
Taking for $j$ the smallest possible value, $j=1/2$, we find that the upper
limit for the number of magnetic impurity atoms is $N_{imp}=0.0016\;N_{A}$,
where $N_{A}$ is Avogadro number. Thus, the average concentration of the 
unspecified magnetic impurities is about $0.1$ at\%, 
which is at least one order of magnitude less than the lowest Ce 
concentration we measure.

\begin{figure}[tbp]
\epsfxsize=3.0in \epsffile{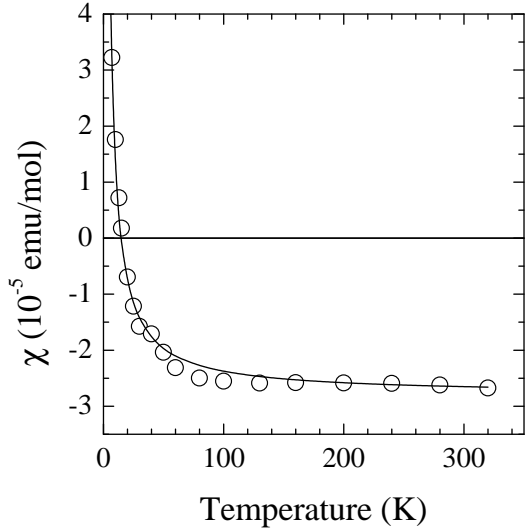}
                                                      \label{fig_matrix}
\caption
{The susceptibility of the LaCu$_{2.05}$Si$_{2}$ matrix 
$\chi (0,T)$ plotted versus temperature (circles). 
The full line is the least-squares fit by Eq.\ (\ref{x=0_background}).
The low-temperature Curie-Weiss upturn is due to magnetic contamination.
Note the difference in the vertical scales used here and in Fig.\ 2.}
\end{figure}

Using Eqs.\ (\ref{sus_x&T}) and (\ref{x=0_background})
we find that $[\chi (x,T)-\chi(0,T)]/x$ defines the same curve
for all the samples. This unique curve defines the impurity susceptibility 
$\chi _{ion}(T)$ up to a constant shift $\left[ \partial \chi
_{B}(x)/\partial x\right]_{x=0}$. However, the two terms in the
braces in Eq.\ (\ref{sus_x&T}) can not be obtained by independent
measurements, and we replace $\chi _{ion}(T)$ in Eq.\ (\ref
{sus_x&T}) with a theoretical expression (for details see below) and find
that the experimental data can be fitted by taking 
$\left[ \partial \chi_{B}/\partial x\right]_{x=0}
=
10^{-4}$($emu/mol$ Ce). 
The same value of this constant shift is obtained by using for the
high-temperature single-ion susceptibility in Eq.\ (\ref{sus_x&T}) the
expression $\chi _{ion}(T)\sim 1/T$, and making the least-squares fit
through the data above 250 K. The comparison of $\chi _{ion}(T)$ and 
$\partial \chi_{B}(x)/\partial x$ 
shows that the susceptibility contribution due to Ce ions is
at least one order of magnitude larger than the background contribution, so
that a possible small error in $\left[ \partial \chi _{B}(x)/\partial
x\right] _{x=0}$ does not influence our conclusions in an essential way. 

\begin{figure}[tbp]
\epsfxsize=3.0in \epsffile{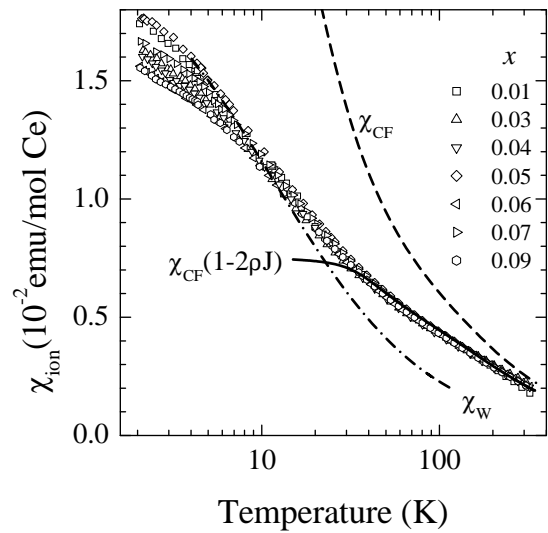}
\caption
{The average single-ion susceptibility of Ce$_{x}$La$_{1-x}$Cu$_{2.05}$Si$_{2}$
defined as $\chi_{ion}(T)=[\chi(x,T)-\chi(0,T)]/x - 1\cdot 10^{-4}$($emu/mol$ Ce),
plotted versus temperature for $x\leq 0.09$ samples. The full line shows the scaling
result, the dashed line shows the CF result and the dashed-dotted 
line shows the exact solution for the spin-1/2 Kondo model 
in the local moment regime. 
                             \label{fig_sus_ion}
}
\end{figure}

The average susceptibility of a single Ce ion in dilute
Ce$_{x}$La$_{1-x}$Cu$_{2.05}$Si$_{2}$ samples is defined by the expression 
\begin{equation}
                             \label{sus3}
\chi_{ion}(T)
=
\frac{\chi(x,T)-\chi(0,T)}{x} - 10^{-4}(emu/mol \;Ce),
\end{equation}
which is shown in Fig.\ref{fig_sus_ion} as a function of temperature 
for various values of $x$. 
The response of Ce ions is nearly the same for all 
the compounds up to $x\simeq 0.09$, and it begins to deviate 
from the $x$-independent form for $x\geq 0.1$.\cite{aviani.99}  
The scatter of the data at low temperatures is about the same 
as the scatter shown in Fig.\ 1, and can be eliminated by plotting 
$\chi_{ion}$ on a universal temperature scale $T/T_K(x)$, 
with $T_K(x)= 8.5\pm1$ K. 
However, in our dilute samples, the lack of any systematic behavior 
of $T_K(x)$ as a function of Ce--doping does not allow us to explain 
the observed fluctuations in terms of the chemical pressure effects 
induced by Ce doping. As discussed in the previous section, 
the random variation of $T_K(x)$ can be associated with the local 
fluctuations in Cu stoichiometry. 

\begin{figure}[tbp]
\epsfxsize=3.0in \epsffile{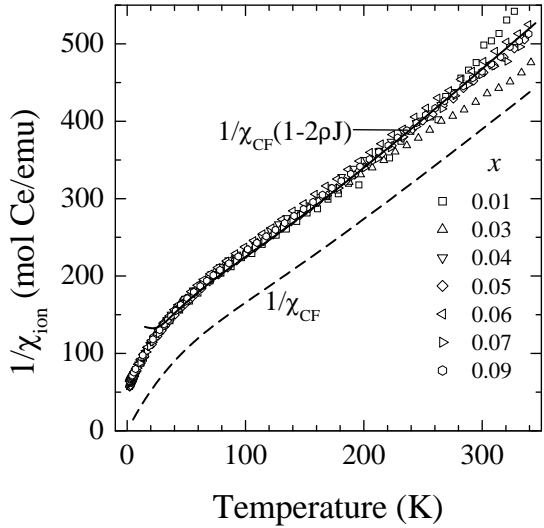}
\caption
{The inverse of the average single-ion susceptibility 
shown as a function of temperature for Ce$_{x}$La$_{1-x}$Cu$_{2.05}$Si$_{2}$  
samples with $x\leq 0.09$. 
Dashed line is the CF result, and the full 
line is the scaling result.
                                                \label{fig_ion_CW}
}
\end{figure}

The anomalous behavior of the molar susceptibility of Ce in 
Ce$_{x}$La$_{1-x}$Cu$_{2.05}$Si$_{2}$ becomes most transparent if one 
compares $\chi_{ion}(T)$ with the susceptibility of a 
crystalographically equivalent but unhybridized $4f^1$ state, $\chi_{CF}(T)$, 
which is calculated in the Appendix. In Fig.\ref{fig_sus_ion} we plot the susceptibilities 
versus $\log T$, and in Fig.\ref{fig_ion_CW} we use the Curie-Weiss plot and  
show the inverse of the susceptibilities as a function of $T$. 
The symbols represent the experimental data, while the dashed line, the full line 
and the dashed-dotted line represent the CF, the scaling, and the 
Wilson's result, respectively. 

The high-temperature susceptibility follows the Curie-Weiss law, 
$1/\chi_{ion}(T)\simeq (T-\Theta )/C$, 
where 
$C=N_{A}\mu_{\mathrm eff}^{2}/3k_{B}
\simeq 
0.125\mu_{\mathrm eff}^{2}/\mu_{B}^{2}
\simeq 0.8$ 
K $emu/mol$ Ce,   
which is close to the CF result. 
The antiferromagnetic Weiss temperature is 
$\Theta \simeq -100 $ K. 
The low-temperature susceptibility data can also be represented by 
a Curie-Weiss law with $C=0.3$ K $emu/mol$ Ce and  
$\Theta \simeq -15$ K. This value of $C$ leads to 
$\mu_{\mathrm eff} \simeq 1.6 \mu_{B}$ which 
is close to the average value obtained for the lowest CF doublet
(see the Appendix).

The analysis of the single-impurity data in 
Figs.\ \ref{fig_sus_ion} and \ref{fig_ion_CW} 
shows clearly that the CF theory, which neglects the coupling 
of $f$  electrons to conduction states, fails to explain the response 
of the $4f^1$ ions to an external magnetic field. 
The correct description is obtained by considering not only 
the thermodynamic fluctuations but taking also into account 
the quantum fluctuations, due to Kondo effect.
Before evaluating these effects in more detail,  
we discuss the torque measurements which provide 
the magnetic anisotropy of Ce ions and allow us to obtain the full 
susceptibility tensor. 

\subsubsection{The susceptibility anisotropy \label{experimental_anisotropy}}
The intrinsic anisotropy of polycrystalline Ce$_{x}$La$_{1-x}$Cu$_{2.05}$Si$_{2}$ 
samples with clear preferential orientation was measured on 
a high-precision torque magnetometer. In the experiment 
the sample is attached to a thin vertical quartz fiber 
and a homogeneous magnetic field ${\bf H}$ is applied in the horizontal $xy$ plane, 
as shown in Fig.\ 7. Thus, only the $z$ component 
of the torque ${\bf \Gamma}$ is measured. 
\begin{figure}[tbp]
                                               \label{fig_setup}
\epsfxsize=3.0in \epsffile{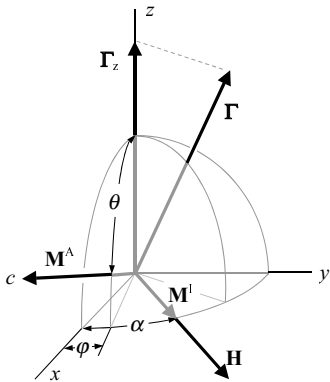}
\caption{The torque on a uniaxial crystallite, 
oriented along $c$ direction, 
induced by the external magnetic field {\bf H} 
in the $xy$ plane.}
\end{figure}
The experimental values for the susceptibility anisotropy of a given 
sample $\Delta\chi^{exp}$ are obtained from the ratio $\Gamma_z/H^2$, 
which is measured by the following procedure.  
For an arbitrary sample orientation, the magnetic field is rotated 
in the $xy$ plane by an angle $\alpha$ varying from 0$^{0}$ to 180$^{0}$.
Measuring the angular dependence of the torque curve at 300 K, 77 K 
and 2 K, in a magnetic field of 8 kOe, we find that $\Gamma_z(\alpha)$ is a 
sinusoidal function of a period $\pi$, with zeros which are 
independent of temperature. This shows that the magnetization of 
the sample is proportional to the applied field. 
The amplitude $\Gamma^{(h)}_z$ of the sine torque curve is then measured as a 
function of temperature. 
Next, the sample is rotated by 90$^{0}$ around the $y$ axis, and 
the amplitude $\Gamma^{(v)}_z$ is measured as a function of temperature. 
The experimental anisotropy of a given sample is defined as 
$\Delta\chi^{exp}_i=2\Gamma_z^{(i)}/H^2 n$ $[emu/mol]$, 
where  $n$ is the number of moles 
and the index $i=h,v$ denotes the two orientations of the sample. 
The data obtained in this way are shown in Fig.\ 8
where $\Delta\chi^{exp}_i(x,T)$ is plotted versus temperature 
for $x=0.01, 0.02, 0.03$, and $0.06$. 
All of the curves in Fig.\ 8 show similar temperature 
dependence, with the extremum point at 48 K. 
The torque signal is quite strong due to a large anisotropy of the Ce 
moment and because the measurements are performed on samples 
with a rather high degree of preferential alignment of the crystallites within 
the polycrystal. 
\begin{figure}[tbp]
                                                \label{fig_torque}  
\epsfxsize=3.0in \epsffile{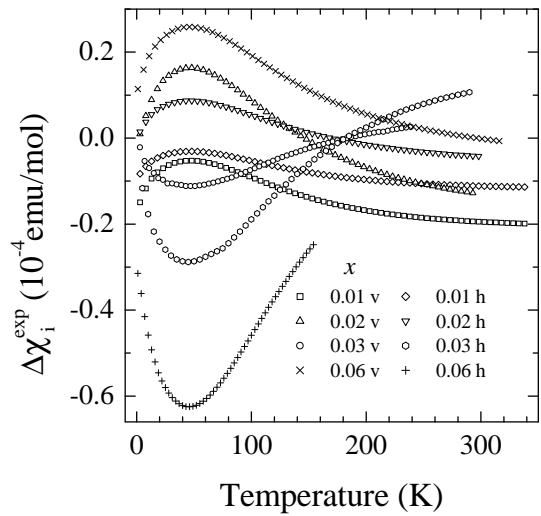}
\caption
{The susceptibility anisotropy of
  Ce$_{x}$La$_{1-x}$Cu$_{2.05}$Si$_{2}$ is shown 
  as a function of temperature for
  $x\leq 0.06$ samples.  
  The data corresponding to two different orientations 
  of the sample with respect to the magnetic field are denoted 
  by $h$ and $v$, respectively.}
\end{figure}

To analyze the torque data shown in Fig.\ 8
we write the $z$ component of ${\bf \Gamma}$ as 
\begin{equation}
                                              \label{torque_z}
\Gamma_{z}
=
V ({\bf M}\times {\bf H})_{z}, 
\end{equation}
where $V$ is the volume of the sample and 
{\bf M} is the magnetization induced by {\bf H}. 
For a single tetragonal crystallite, the induced magnetization 
has an isotropic and an anisotropic component, 
i.e., it can be written as
\begin{equation}
{\bf M}
=
{\bf M}^{I}+{\bf M}^{A}, 
\label{M}
\end{equation}
where the isotropic component {\bf M}$^{I}$ is directed along {\bf H}, 
while the anisotropic one {\bf M}$^{A}$ is directed along the 
$c$ axis of the crystal. Obviously, ${\bf M}^{I}=\chi^{ab}{\bf H}$ and 
${\bf M}^{A}=(\chi^c-\chi^{ab}){\bf H}_c$,
where $\chi^c$ is the susceptibility along the $c$ axis, 
$\chi^{ab}$ is the susceptibility in the $ab$ plane, and ${\bf H}_c$ 
is the component of the magnetic field along the $c$ direction. 
Using Eq.(\ref{torque_z}), we find that 
the torque on a crystallite with the $c$ axis directed along 
$(\theta, \varphi)$, 
and induced by the magnetic field in the $xy$ plane, 
${\bf H}=H(\cos\alpha,\sin\alpha,0)$, 
is given by the expression  (see also Fig.\ 7)
\begin{equation}
                                          \label{torque}
\Gamma_{z}
=
\frac{1}{2}n
H^{2}\Delta\chi 
\sin ^{2}(\theta)\sin\left[2(\alpha -\varphi )\right],  
\end{equation} 
where $\Delta\chi$ is now measured in $emu/mol$, 
i.e., stands for $(V/n)\Delta\chi$. 
Thus, the temperature dependence of the torque is given,
up to a numerical constant, by the intrinsic susceptibility anisotropy 
$\Delta\chi=\chi^c-\chi^{ab}$, regardless of the orientation of 
the crystallite.

In a polycrystalline sample, we are dealing with some volume distribution 
$V_{i}=V_{i}(\theta _{i},\varphi _{i})$ of the crystallites 
with respect to their orientations. The total torque on a  sample 
with a given concentration of Ce ions is the sum of all the separate 
contributions and can be written as 
\begin{equation}
                          \label{torque_poly}
\Gamma^x_{z}
=
\frac{1}{2}n
H^{2}\gamma_x \Delta \chi 
\sin \left[2(\alpha -\bar{\varphi}_x)\right],   
\end{equation}
where we introduced the alignment factor $\gamma_x $ 
$(0\leq \gamma_x \leq 1)$ and the phase shift $\bar{\varphi}_x$. 
In our experiment the magnet is rotated to a position 
$\alpha -\bar{\varphi}_x=\pm\pi/4$ which maximizes the torque. 
Thus, up to a geometrical factor $\gamma_x$, the temperature dependence 
of the torque follows from the intrinsic anisotropy of the sample. 
The alignment factor $\gamma_x$ depends on the distribution of 
the polycrystallites in the sample, and is given by 
$\gamma_x=\sqrt{A_x^{2}+B_x^{2}}$,  where
$A_x=\frac{1}{V}\sum_{i}V_{i}\sin ^{2}\theta _{i}\cos 2\varphi _{i}$ 
and 
$B_x=\frac{1}{V}\sum_{i}V_{i}\sin ^{2}\theta _{i}\sin 2\varphi _{i}$. 
To find the single-ion anisotropy we have to estimate $\gamma_x$ 
and the intrinsic anisotropy of the metallic matrix. 

The data analysis shows that the experimental anisotropy 
$\Delta\chi^{exp}(x,T)=\gamma_x \Delta\chi$ follows a similar 
temperature dependence in all the samples.
This is shown in Fig.\ 9, where $\Delta\chi_i^{exp}(x,T)$  
is plotted as a function of $\Delta \chi_v^{exp}(x_1,T)$, 
with temperature as an implicit variable. We recall that index $i=v,h$ 
denotes the orientation of the sample with respect to the applied field. 
The orientation of the reference sample is $v$ and the referent Ce 
concentration is $x_1=0.01$.  
The correlation between the torque data is found to be strictly linear, 
\begin{equation}
                                   \label{anizokor}
\Delta \chi_i^{exp}(x,T)
=
k_x^i \Delta\chi^{exp}_v(x_{1},T)+l_x^i , 
\end{equation}
where $k_x^i$ and $l_x^i$ denote the slope and the intercept 
of the straight lines  respectively.
The numerical values of $k_x^i$ and $l_x^i$ 
for each sample are given in Table I. 
As shown below, the linear relation (\ref{anizokor}) 
proves essential for obtaining the single-ion contribution 
from the experimental anisotropy data. 
\begin{figure}[tbp]
\epsfxsize=3.0in \epsffile{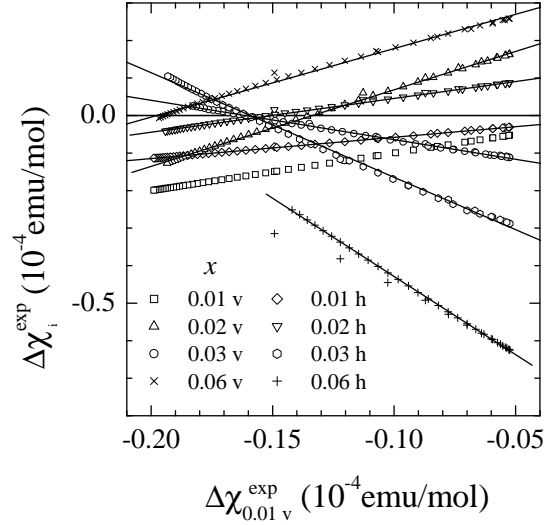}
                                         \label{fig_anizo_correl}   
\caption
{$\Delta \chi_i^{exp} (x,T)$ of Ce$_{x}$La$_{1-x}$Cu$_{2.05}$Si$_{2}$ 
samples is plotted versus $\Delta \chi_v^{exp} (x_1,T)$. 
The data are given for two different orientations ($i=h,v$) with 
respect to {\bf H}. The concentration of the Ce ions in the referent 
sample is $x_1=0.01$. 
The full lines are the linear extrapolations which define the 
coefficients $k_x^i$ and $l_x^i$.}
\end{figure}
In analogy with the analysis of the susceptibility data 
we assume that the susceptibility anisotropy of a given crystallite 
has a single-ion contribution, 
$\Delta\chi_{ion}(T)=\chi_{ion}^c-\chi_{ion}^{ab}$, 
and a van Vleck contribution due to the matrix, 
$\Delta \chi_{B}=\chi_{B}^c-\chi_{B}^{ab}$. 
The unspecified magnetic impurities which contaminate our samples, 
and which are described by Eq.\ (\ref{x=0_background}), 
are assumed to be isotropic, because the the torque shown 
in Fig.\ 8 has no Curie-like upturn. 
Thus, we write the susceptibility anisotropy of a given sample as 
\begin{equation}
                                     \label{anizo1}  
\Delta\chi_i^{exp}(x,T)
=\gamma _{x}^{i}\left[ x\,\Delta\chi_{ion}(T)
+\Delta\chi_{B}(x)\right],  
\end{equation}
where the temperature dependence of $\Delta\chi_{B}$ 
is neglected. 
In the dilute limit, we approximate the van Vleck anisotropy 
by a linear expression 
$\Delta\chi_{B}(x)
\simeq 
\Delta\chi_0
+
x\left[
       \frac{\partial\Delta\chi_{B}(x)}{\partial x}
\right]_{x=0}$,   
and use Eqs.\ (\ref{anizokor}) and (\ref{anizo1}) to derive an approximate 
relation between the experimentally determined quantities 
\begin{equation}
                                    \label{anizo-omjer1}
\frac{l_x^i x}{k_x^i (x-x_{1})}
=
-\gamma^v_{1}\Delta \chi_0,
\end{equation}
where $\gamma^v_{1}$ is the alignement factor of the referent sample. 
The experimental values obtained from Fig.\ 9 
indicate that the ratio in Eq.\ (\ref{anizo-omjer1}) 
is indeed very nearly sample-independent, and given by 
$\gamma^v_{1}\Delta \chi_0\simeq -2.4\cdot 10^{-5}$$[emu/mol]$. 
Using this value, and Eqs.\ (\ref{anizokor}) and (\ref{anizo1}), 
we can express the product ${\gamma^v_1}\Delta \,\chi_{ion}(T)$ 
in terms of experimentally determined quantities, \cite{anizo_corr} 
\begin{equation}
                                      \label{anizo-extract}
{\gamma_1^v}\Delta \,\chi_{ion}(T)
=
   \left[ 
     \frac{\Delta\chi_i^{exp}(x,T)-l_x^i}{x_1 k_x^i}
-
     \frac{\gamma^v_1 \Delta\chi_0}{x_1}
   \right].      
\end{equation}
Considering ${\gamma^v_1}\Delta \,\chi_{ion}(T)$ as a function of 
temperature one obtains the same function for all the samples. 
The value of $\gamma^v_1$ is obtained by fitting the universal 
curve (\ref{anizo-extract}) by the theoretical result for the 
Coqblin-Schrieffer model, with the same parameters as in the 
previous section. This gives $\gamma^v_{1}=0.55$.  
\begin{figure}[tbp]
\epsfxsize=3.0in \epsffile{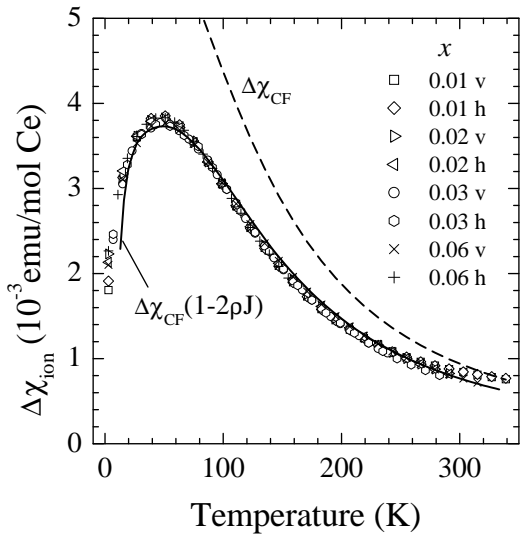}
\caption
{The single-ion susceptibility anisotropy 
$\Delta\chi_{ion}(T)$ defined by Eq.\ (\ref{anizo-extract}) 
plotted versus temperature for all the Ce$_{x}$La$_{1-x}$Cu$_{2.05}$Si$_{2}$
samples with $x\leq 0.06$. The full line is the scaling result, and the dashed line is the 
CF result.
                                                    \label{Aniz_Ce_ion.eps}   
}
\end{figure}
The single-ion anisotropy $\Delta\chi_{ion}(T)$, defined in the  
above described way, is plotted as a function of temperature in Fig.\ 10, 
together with the scaling result (full line) and the CF 
anisotropy (dashed line). 
The data show clearly that the magnetic anisotropy of $4f^1$ 
ions in the tetragonal Ce$_{x}$La$_{1-x}$Cu$_{2.05}$Si$_{2}$ crystals 
can not be explained by the CF theory, which considers only 
the thermal fluctuations among various CF states. The correct 
description is obtained by including the quantum fluctuations, 
and is provided here by the scaling solution of the 
Coqblin-Schrieffer model.
\subsection{Theoretical analysis \label{scaling} }

The properties of Ce ions in dilute Ce$_{x}$La$_{1-x}$Cu$_{2.05}$Si$_{2}$ samples 
are modeled by the Coqblin-Schrieffer Hamiltonian with CF splitting, 
\begin{equation}  \label{A}
H_K=H_{CF}+H_0+H_J,
\end{equation}
where $H_{CF}$ describes the $f$  states in the tetragonal 
CF (see the Appendix), 
$H_0$ describes the conduction band of width $2D_0$, with a constant 
density of states $\rho $, 
and $H_J$ defines the exchange coupling between the $4f^1$ states and 
the band electrons. \cite{coqblin.69,coqblin.72} 

The high-temperature properties are calculated by the poor man's scaling, 
\cite{anderson.70,yamada.84,hanzawa.85}  
which provides the renormalized coupling by reducing the 
conduction electron cutoff from $D_0$ to $D$, and simultaneously rescaling 
the coupling constant from $J_0$ to $J(D)$, 
so as to keep the low-energy excitations of the total system unchanged.  
Assuming the isotropic exchange, this leads to \cite{yamada.84,hanzawa.85}
\begin{equation}
                                    \label{hanzawa_3}
\rho J
\exp \left( -\frac{1}{\rho J}\right)
=
\left(\frac{k_B T_{K}}{D}\right)^{m}
\left(\frac{k_B T_{K}+
\Delta }{D+\Delta }\right)^{M},
\end{equation}
where $J$ is the absolute value of the antiferromagnetic coupling constant,
$m=2$ and $M=4$ are the degeneracies of the lower and the upper CF state,  
respectively, $\Delta$ is the CF splitting, and $T_K$ is the Kondo temperature. 
The renormalized coupling constant $J(D)$ defines an effective model with cutoff $D$. 
The Kondo temperature is given by Eq.\ (\ref{hanzawa_3}) with $J$ and $D$ 
replaced by the initial values $J_0$ and $D_0$, respectively
(for details see \cite{yamada.84,hanzawa.85}).  
The scaling law (\ref{hanzawa_3}) applies provided the 
renormalized coupling is small, i.e., $\rho J\ll 1$ and  $D \gg T_K$.  

The scaling trajectory described by Eq.(\ref{hanzawa_3}) has two asymptotic regimes. 
The high-temperature asymptote, $J_H(T)$, is obtained by neglecting 
the CF splitting, and is given by, 
\begin{equation}
                                    \label{T_K^H}
\rho J_H
\exp \left( -\frac{1}{\rho J_H}\right)
=
\left(\frac{k_B T^H_K}{D}\right)^{N}, 
\end{equation} 
where, $T^H_K$ is the high-temperature Kondo scale, defined by 
Eq.(\ref{T_K^H}) with $J_0$ and $D_0$ replacing $J_H$ and $D$, respectively. 

The low-temperature asymptote $J_L(D)$ is obtained by neglecting 
in Eq.(\ref{hanzawa_3}) the effects of the excited CF levels, 
i.e., by neglecting $T_K$ and $D$ with respect to $\Delta$. 
This identifies $J_L(D)$ as the scaling trajectory of an effective doublet 
with Kondo scale $T_K$, and the effective coupling, 
\begin{equation}
                                  \label{T_K^L}
\rho J_L
\exp \left( -\frac{1}{\rho J_L}\right)
=
\left(\frac{k_B T_{K}}{D}\right)^{m}.    
\end{equation}
Eqs.(\ref{hanzawa_3} - \ref{T_K^L}),
show that the $f$ level behaves at high temperatures as an effective sextet 
with Kondo scale $T_K^H$, and at low temperatures as an effective 
doublet with Kondo resonance $T_K$. 
The two Kondo scales satisfy the relation, 
\begin{equation}
                                    \label{T_K^H-T_K}
(k_B T_K^H)^{m+M}=(k_B T_K)^m\Delta^M.
\end{equation}
The scaling result given by Eq.(\ref{hanzawa_3}) shows that we can remove the CF 
splitting from the asymptotic considerations at high and low 
temperatures, provided we renormalize the effective coupling 
constant in such a way that the low-energy properties of the effective 
models and the Coqblin-Schrieffer model are the same. 
Note, the $T_K$ of the effective doublet is much higher than the Kondo 
scale of the  Coqblin-Schrieffer model calculated in the limit 
$\Delta\gg D_0$, with the initial condition $J_L(D_0)=J_0$. 

The anisotropic susceptibility tensor of the Coqblin-Scrieffer 
model with CF splitting is obtained by applying the linear 
response theory, and calculating the response functions by 
lowest-order perturbation theory with renormalized coupling 
constants\cite{kww.80,hewson.93,chen.92,yosida.95}. 
The effective Coqblin-Schrieffer model at a given temperature T 
is obtained by reducing the cutoff from $D_0$ to $D=A \, k_B T$, 
and renormalizing $J_0$ to $J(T)=J(T_K/A T,\Delta/A k_B T)$ 
according to Eq.\ (\ref{hanzawa_3}).
Here,  A is a numerical constant of the order of unity. 
Thus, we obtain the susceptibility  
\begin{equation}                  
                                  \label{chi_J(T)}
\chi^{\alpha}_J(T)
=
\chi_{CF}^{\alpha}(T)
[1-2\rho J(T)],  
\end{equation}
where $\chi_{CF}^{\alpha}(T)$ is the $\alpha$ component of 
the anisotropic CF susceptibility calculated in the Appendix.
The result \ (\ref{chi_J(T)}) shows that the exchange interaction 
reduces the CF susceptibility, and the poor man's scaling  
gives the reduction factor as $F(T,T_K,\Delta)=[1-2\rho J(T)]$. 
Fitting the experimental results above 30 K with the renormalized 
susceptibility given by Eq.(\ref{chi_J(T)}),  
where $J(T)$ is given by Eq.(\ref{hanzawa_3}), 
we obtain an implicit relation between $T_K$ and the cutoff constant A.  
The analysis of the Ce$_x$La$_{1-x}$Cu$_{2.05}$Si$_2$ data above 30 K 
shows that we van write $T_K\simeq\alpha A +\beta$, 
where $\alpha=3$ K and $\beta=-1$ K. 
The physically acceptable range of the cutoffs is from $k_B T$ to $4k_B T$, 
which gives $T_K$ between 2 K and 11 K, and $T_K^H$ between 60 and 110 K. 
However, the analysis of the low--temperature data (see below) leads to 
the values $T_K=8.5$ and $A=3$.
The theoretical results obtained in such a way are shown in 
Figs.\ \ref{fig_sus_ion}, \ref{fig_ion_CW}, \ref{Aniz_Ce_ion.eps}, 
\ref{Sus_abc_CW.eps}, and \ref{reduction_factor.eps} as a full line. 

The experimental data are described by the scaling theory down to 30 K. 
At lower temperatures there is a discrepancy, which is not really surprising, 
since $J(T)$ in Eq.(\ref{chi_J(T)}) grows rapidly for $T\ll T_K^H$ 
and at about 30 K the coupling given by Eq.(\ref{hanzawa_3}) 
becomes too large for the renormalized perturbation expansion to be valid. 
Note, in the absence of the CF splitting the perturbation 
theory for the $N$-fold degenerate $f$  level breaks down at around 
$T\simeq T_K^H$. 
The CF splitting reduces the degeneracy of the ground state and 
extends the validity of the perturbation expansion 
down to about $T\simeq A T_K$, where $T_K$ is the Kondo temperature of 
the effective doublet, and $T_K\ll T_K^H$. 

Below 20 K, however, the coupling constants defined by Eqs.(\ref{hanzawa_3}) 
and (\ref{T_K^L}) satisfy $J(T)\simeq J_L(T)$, such that we can consider 
the $f$ state as an effective doublet. 
The average moment of the CF ground state, calculated by the CF theory, 
is $\mu=1.62 \mu_B$, which is not too different from the free spin-1/2 
moment, $\sqrt{3}\mu_B$.
Thus, to discuss the low-temperature data, for $T\leq T_K$ we neglect 
the small anisotropy of the CF ground state and approximate 
the CF split multiplet by an effective doublet, and replace the 
Coqblin-Schrieffer model by an effective spin-1/2 Kondo model. 
We relate the two models by setting the Kondo temperature
of the spin-1/2 Kondo problem to $T_K$. In such way the parameter 
space of the low--temeperature model is completly determined. 
The susceptibility of the Kondo model in the local moment (LM) regime, 
i.e., for $T_K/2\leq T\leq 15 T_K$, is given by the numerical 
renormalization group result\cite{wilson.75} 
\begin{equation}                  
                                  \label{wilson}
\chi_W
=
\frac{0.68 C_0}{T+\sqrt{2}T_K} , 
\end{equation}
where $C_0={N_A(g\mu_B)^2S(S+1)}/{3k_B}$, i.e., $C_0=0.375$ $emu/mol$.
Using $\chi_W$ and $T_K=8.5$ K we obtain the curve shown in Figs.\ 5 
and 12. In the low-temperature LM regime, which sets in for diluite 
Ce$_{x}$La$_{1-x}$Cu$_{2.05}$Si$_{2}$ alloys between 4 K and 30 K, 
we find that the calculated susceptibility is very close to the 
experimental values. 
Thus, by demanding that the low-- and the 
high--temperature models have the same Kondo temperature we restricted 
the high--temperature cutoff constant to $A=3$. 

\subsection{Discussion of the experimental results and 
conclusion\label{discussion}} 
\begin{figure}[tbp]
\epsfxsize=3.0in \epsffile{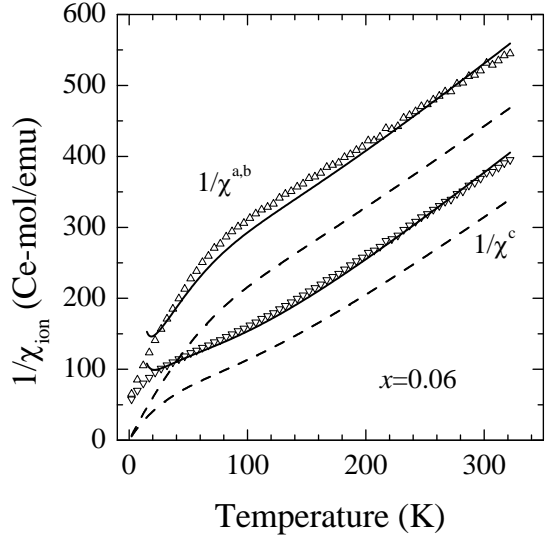} 
\caption
{The single-ion susceptibility for finite Ce concentration 
shown on a Curie-Weiss plot as a function of temperature for 
Ce$_{0.06}$La$_{0.94}$Cu$_{2.05}$Si$_{2}$. 
Dashed line is the CF result  $1/\chi^{\alpha}_{CF}(T)$,  
and full line is the scaling result $1/\chi^{\alpha}_{J}(T)$. 
The calculations are performed using $T_K=8.5$ K,  $T_K^H=100$ K, 
and $\Delta= 330$ K.
                                              \label{Sus_abc_CW.eps}
}
\end{figure}

The anisotropic susceptibility of a single Ce ion is found
by systematic measurements of dilute Ce$_{x}$La$_{1-x}$Cu$_{2.05}$Si$_{2}$
alloys on the Faraday balance and the torque magnetometer, 
and by carefully subtracting the background. 
The experimental results are explained by the Coqblin-Schrieffer model 
with the CF doublet-quartet splitting $\Delta=330$ K 
and with the exchange scattering such that $T_K=8.5\,$K. 
The average value of the calculated susceptibility tensor, 
$\chi_J(T)= [2\chi^{ab}_J(T)+ \chi^{c}_J(T)]/3$, 
is shown in Figs.\ 5 and 6, together with the experimental data. 
The corresponding result for the susceptibility anisotropy 
$\Delta\chi_J(T)$ is compared with the torque measurements 
in Fig.\ \ref{Aniz_Ce_ion.eps}. 
Combining the Faraday balance and the torque data we find the 
single-impurity response along the principal axes, which is shown 
in Fig.\ref{Sus_abc_CW.eps} for the $x=0.06$ sample, together 
with the scaling result (full lines) and the CF theory (dashed lines).
The experimental data for other samples are about the same.
The principal-axes susceptibilities shown in Fig.\ \ref{Sus_abc_CW.eps} 
follow between 100 K and 350 K an anisotropic Curie-Weiss law, 
\begin{equation}
\chi^{\alpha}_{ion}
=
\frac{C^\alpha}{T-\Theta^\alpha}
                                   \label{1/chi(T)}, 
\end{equation}
such that the slope of $1/\chi^{\alpha}_{ion}(T)$ and $1/\chi^{\alpha}_{CF}(T)$
is about the same. Thus, the high-temperature data can be discussed in 
terms of an anisotropic local moment which is close to the CF value. 
However, the response of the $f$ state is reduced with respect to the 
CF value due to the temperature dependent exchange coupling to the 
conduction band. The lowest order perturbation theory gives the reduction 
factor as $F(T/T_K^H)=1-2\rho J_H(T/T_K^H)$, where we assumed an isotropic 
exchange coupling.
The Ce impurity behaves in this high-temperature LM regime 
as a sextet split by the tetragonal CF, and with the relevant Kondo 
scale $T_K^H=100$ K. 

Below $T\simeq T_K^H$, we observe the crossover from the high-temperature 
LM regime to an effective twofold degenerate low-temperature LM regime. 
Surprisingly, the behavior of the system at the crossover is still 
rather well described by scaling, 
and the single-ion response can be discussed in terms of the reduced 
CF susceptibility. 

\begin{figure}[tbp]
\epsfxsize=3.0in \epsffile{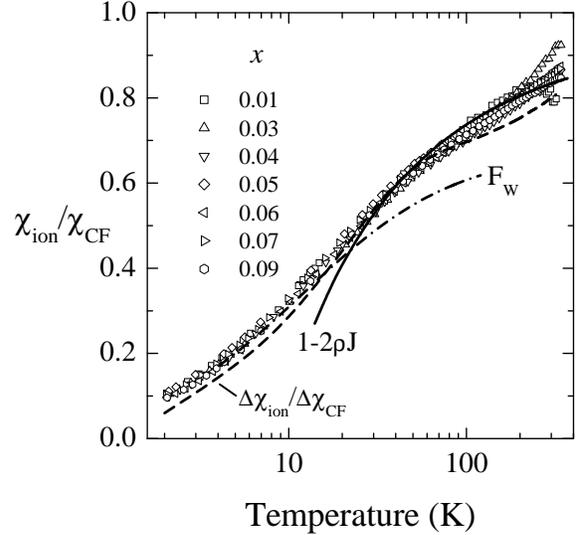}  
\caption
{Temperature dependence of the reduction factor of the susceptibility,
$F(T)=\chi_{ion}(T)/\chi_{CF}(T)$ for Ce$_{x}$La$_{1-x}$Cu$_{2.05}$Si$_{2}$
with $x\leq 0.09$. The symbols are obtained from the susceptibility data,
the dashed line is obtained from the anisotropy data and the solid line is
the scaling result. The dashed-dotted line shows the reduction factor 
for the spin-1/2 Kondo model. 
                                \label{reduction_factor.eps}
}
\end{figure}

At low temperatures, $T \leq A\ T_K$, the renormalized coupling becomes too 
large for the lowest-order renormalized perturbation expansion to be valid. 
Since we are not aware of any accurate theoretical results for the response 
of Coqblin-Schrieffer model with CF splitting that we can use close to $T_K$, 
we estimate the reduction factor $F(T)$ directly from the experiments, 
assuming that the exchange is isotropic 
and that the susceptibility retains the factorized form 
(\ref{chi_J(T)}) down to lowest temperatures. 
The Faraday balance data give $F(T)=\chi_{ion}(T)/\chi_{CF}(T)$ 
and the torque data give $F(T)=\Delta\chi_{ion}(T)/\Delta\chi_{CF}(T)$, 
which we plot in Fig.\ref{reduction_factor.eps}, together with the 
high-temperature scaling result $\chi_J/\chi_{CF}=[1-2\rho J(T)]$ 
and the Wilson's result $F_W=T\chi_W/C_0$. 
The reduction factors obtained from the average susceptibility and 
the anisotropy data are very similar, and the high--temperature 
data are rather well described by the poor man's scaling. 
At low temperatures, where the crystal field calculations 
lead to an effective doublet with 
$\chi_{CF}\simeq N_A \mu_{\mathrm eff}^2/3k_B T$ 
and 
$\mu_{\mathrm eff}\simeq 1.621 \mu_B$,
the  experimental reduction factor comes very close to the universal 
curve obtained for the isotropic spin-1/2 Kondo model.\cite{wilson.75}  

We mention, for completeness, that the magnetic anomalies discussed 
here are accompanied by the usual Kondo anomalies in the electric 
resistivity and the thermoelectric power data. \cite{ocko.99} 
All the samples used for the susceptibility  measurements have  
a clear Kondo minimum in the bare resistivity, as well as 
a very large thermoelectric power with two peaks, which is typical 
of CF split Kondo ions.\cite{zlatic.93} The low-temperature 
peak is at about 10 K, and the high temperature one at about 120 K, 
as expected in a system described by the Coqblin-Schrieffer model 
with the Kondo scales, $T_K\simeq 8.5$ K and $T_K^H\simeq 100$ K. 

In summary, the magnetic susceptibility of a dilute Ce ions embedded 
in the tetragonal metallic host has been obtained by careful data 
analysis. The samples used in our studies have a negligibly 
small Ce--Ce interaction. The changes in the matrix induced by the 
doping are also found to be very small. 
The behavior of the Ce$_{x}$La$_{1-x}$Cu$_{2.05}$Si$_{2}$ compounds 
with less than 9 at\% of Ce is well described by the Coqblin-Schrieffer 
model of a $f$ state comprising a ground state doublet and a pseudoquartet  
split by $\Delta\simeq 330$ K. 
The quantum fluctuations due to the exchange coupling between the $f$ state 
and the conduction band can be described by the poor man's scaling, which 
explains the high-temperature data and the crossover from the 
high-temperature LM regime to a twofold degenerate low-temperature LM regime, 
that takes place at about $T^H_K\simeq 100$ K. The lowest-order 
perturbation expansion, based on the scaling solution of the CF split model,  
breaks down below 30 K. However, at such low temperatures, the effect 
of the excited CF states is rather small, and the experimental results 
below 20 K can be described by the exact solution of the spin-1/2 
Kondo model with Kondo scale $T_K=8.5$ K. 
This low--temperature Kondo scale determines completly the cutoff 
constant used in the high--temperature scaling. 

\subsection{Acknowledgments}
We acknowledge the useful comments from B. Horvati\'c.  
The financial support from the Alexander von Humboldt Foundation 
to VZ is gratefully acknowledged. 

\subsection{Appendix                    \label{appendix_CF}} 
A Ce$^{3+}$ ion in a tetragonal crystal field is described by the 
Hamiltonian\cite{Hutchings.64}
\begin{equation}
H_{CF}=B_{2}^{0}O_{2}^{0}+B_{4}^{0}O_{4}^{0}+B_{4}^{4}O_{4}^{4},
\label{H_CF}
\end{equation}
where $B_{l}^{m}$ are the CF parameters and $O_{l}^{m}$
are the Stevens operators, which are related to the angular momentum
operator $J$ and its components $J_{x}=\frac{1}{2}(J_{+}+J_{-})$, 
$J_{y}=\frac{1}{2i}(J_{+}-J_{-})$, and $J_{z}$, 
as follows: 
\begin{equation}
O_{2}^{0}=3J_{z}^{2}-J(J+I),  \label{O20}
\end{equation}
\[
O_{4}^{0}=35J_{z}^{2}-30J(J+I)J_{z}^{2}+25J_{z}^{2}-6J(J+I)+3J^{2}(J+I)^{2},
\]
\[
O_{4}^{4}=\frac{1}{2}(J_{+}^{4}+J_{-}^{4}).
\]

For small CF this Hamiltonian is a perturbation to the sixfold degenerate $%
4f^{1},$ $j=5/2$ atomic wave functions $\left| \pm \frac{1}{2}\right\rangle
,\left| \pm \frac{3}{2}\right\rangle ,\left| \pm \frac{5}{2}\right\rangle ,$
and it is easily diagonalized. 
The eigenvectors $\left| n\right\rangle$ are given by 
\begin{eqnarray}
                                 \label{H_CFfunkcije}              
\Gamma _{7}^{(1)} 
&:&
\left| \pm 1\right\rangle =\eta \left| \pm 
\frac{5}{2}\right\rangle 
+
\sqrt{1-\eta ^{2}}\left| \mp \frac{3}{2}\right\rangle ,\nonumber\\ 
\hspace{0.5in}
\Gamma _{7}^{(2)} 
&:&
\left| \pm 2\right\rangle =-\sqrt{1-\eta
^{2}}\left| \pm \frac{5}{2}\right\rangle 
+
\eta \left| \mp \frac{3}{2}\right\rangle ,   \\
\Gamma _{6} 
&:&\left| \pm 3\right\rangle = 
\left| \pm \frac{1}{2}\right\rangle ,  \nonumber
\end{eqnarray}
and the energy eigenvalues $\epsilon _{n}$ are
\begin{eqnarray}
                            \label{H_CFenergije}
\epsilon _{1} 
&=&
10B_{2}^{0}
+60B_{4}^{0}
+12\sqrt{5}\frac{\sqrt{1-\eta ^{2}}}{\eta }B_{4}^{4}, \nonumber\\
\epsilon _{2} 
&=&
-2B_{2}^{0}
-180B_{4}^{0}
-12\sqrt{5}\frac{\sqrt{1-\eta^{2}}}{\eta }B_{4}^{4},\\
\epsilon _{3} 
&=&
-8B_{2}^{0}
+120B_{4}^{0}, \nonumber
\end{eqnarray}
where the mixing parameter $\eta$ is defined by the equation 
\begin{equation}
\frac{\eta }{\sqrt{1-\eta ^{2}}}-\frac{\sqrt{1-\eta ^{2}}}{\eta }=\frac{%
B_{2}^{0}+20B_{4}^{0}}{\sqrt{5}B_{4}^{4}}.  \label{H_CFomjer}
\end{equation}

The neutron scattering data\cite{goremychkin.93} indicate 
a $\Gamma_7$ ground state and a CF scheme shown in Fig.\ 13.  
\begin{figure}[tbp]
\epsfxsize=2.0in \epsffile{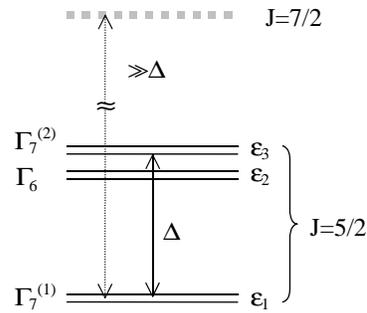}
                                   \label{fig.levels}
\caption
{The CF level for the 
doublet-quartet scheme used 
in the calculations.}
\end{figure}
Since  $\epsilon_{2}\approx \epsilon_{3}$, we use here an 
approximate doublet-quartet scheme. 

The $\alpha$ component of the CF susceptibility of one mole of 
isolated $Ce^{3+}$ ions is given by the van Vleck formula \cite{VanVleck.32}
\begin{eqnarray}
\chi_{CF}^{\alpha } &=&\frac{N_{A}(g_{J}\mu _{B})^{2}}{Z}
\mathop{\displaystyle \sum }
\limits_{n}\left[ \beta \sum_{m,\epsilon _{n}
=\epsilon _{m}}\left| \langle
m\mid J_{\alpha }\mid n\rangle \right| ^{2}e^{-\beta \epsilon _{n}}\right. 
\nonumber \\
&&+\left. \sum_{m,\epsilon _{n}\neq \epsilon _{m}}\left| \langle m\mid
J_{\alpha }\mid n\rangle \right| ^{2}\frac{e^{-\beta \epsilon
_{m}}-e^{-\beta \epsilon _{n}}}{\epsilon _{n}-\epsilon _{m}}\right]
\label{VVsus},
\end{eqnarray}
where $n,m\in \left\{ \pm 1,\pm 2,\pm 3\right\}$, 
$N_{A}$ is the Avogadro number, $\mu _{B}$ the Bohr magneton, $g_{J}$
the Land\'{e} gyromagnetic factor, $Z$ the partition function, 
$\beta=1/k_{B}T$, and 
$\epsilon_{n},\epsilon _{m}$ are given by Eq.\ (\ref{H_CFenergije}). 
The matrix elements $\langle m\mid J_{\alpha }\mid n\rangle$
of the angular momentum $J$ in $\alpha $ direction are taken 
between the CF eigenstates (\ref{H_CFfunkcije}). 

The $n$ summation is performed over all the energy levels, 
while the $m$ summation is performed for degenerate 
$(\epsilon _{n}=\epsilon _{m})$ and 
for non-degenerate $(\epsilon _{n}\neq \epsilon _{m})$ levels separately. 
If the energies are measured relative to $\epsilon _{1}$, we obtain
\begin{eqnarray}
\chi^{c}_{CF} 
&=&
\frac{N_{A}(g_{J}\mu _{B})^{2}}{2\left( 1+2e^{-\beta \Delta
}\right) }\beta \left\{ \left[ 2\left( \frac{3}{2}-4\eta ^{2}\right) ^{2}+%
\frac{1}{2}e^{-\beta \Delta }\right. \right.  \label{susCe_c} \\
&&\left. \left. +2\left( \frac{5}{2}-4\eta ^{2}\right) ^{2}e^{-\beta \Delta
}\right] +64\eta ^{2}\left( 1-\eta ^{2}\right) \frac{1-e^{-\beta \Delta }}{%
\Delta }\right\} ,  \nonumber
\end{eqnarray}
\begin{eqnarray}
\chi^{ab}_{CF}  
&=&\frac{N_{A}(g_{J}\mu _{B})^{2}}
{2\left( 1+2e^{-\beta \Delta}\right)}\beta \left\{ \left[ 10\eta ^{2}\left( 1-\eta ^{2}\right) \left(
1+e^{-\beta \Delta }\right) \right. \right.  \nonumber \\
&&\left. \left. +\frac{9}{2}e^{-\beta \Delta }+8\eta ^{2}\right] +\right. 
\nonumber \\
&&+\left. \left[ 8\left( 1-\eta ^{2}\right) +5\left( 1-2\eta ^{2}\right)
^{2}\right] \frac{1-e^{-\beta \Delta }}{\Delta }\right\}.  \label{susCe_ab}
\end{eqnarray}
The anisotropy $\Delta \chi =$ $\chi ^{c}-\chi ^{ab}$ is 
\begin{eqnarray}
\Delta \chi_{CF}  
&=&
\frac{N_{A}(g_{J}\mu _{B})^{2}\left( 6\eta ^{2}-1\right)}
{2\left( 1+2e^{-\beta \Delta}\right) }
\left\{ \frac{\beta }{2}\left[ 10\eta
^{2}\left( 14\eta ^{2}-17\right) e^{-\beta \Delta }\right] \right.  \nonumber
\\
&&+\left. \left( 14\eta ^{2}-9\right) +\left( 13-14\eta ^{2}\right) \frac{%
1-e^{-\beta \Delta }}{\Delta }\right\} .  \label{anizo_CF}
\end{eqnarray}

In both, the low and the high temperature limits the CF susceptibility in 
$\alpha $ direction can be approximated by the Langevin formula, $\chi
_{CF}^{\alpha }=N_{A}\mu _{\alpha }^{2}/3k_{B}T,$ where $\mu _{\alpha }$ is
the effective magnetic moment. At high temperatures, $T\gg \Delta ,$ the
effective moment is isotropic, 
$\mu _{ab}=\mu _{c}=g_{J}\sqrt{j(j+1)}\mu
_{B}=2.535\mu _{B}$. 

At low temperatures, $T\ll \Delta ,$ the magnetic moment is determined by
the $\Gamma _{7}^{(1)}$ground state doublet moment and it is anisotropic

\begin{eqnarray}
\mu _{c} &=&\sqrt{3}g_{J}\left| \frac{3}{2}-4\eta ^{2}\right| \mu _{B},%
\hspace{0.5cm}  \nonumber \\
\mu _{ab} &=&\sqrt{15}g_{J}\eta \sqrt{1-\eta ^{2}}\mu _{B}.
\label{moment_c_ab}
\end{eqnarray}

The curves shown in the text are calculated with parameters 
$\Delta =330$ K, $\eta =0.816$, and $g_{J}=8/7$. 
The corresponding values for the CF parameters are 
$B_{2}^{0}=-1.01$ meV, $B_{4}^{0}=0.011$ meV, and $B_{4}^{4}=-0.50$ meV.

From those parameters we find $\mu _{c}=1.727\mu _{B}$ and $\mu _{ab}=1.566\mu
_{B}.$ and the effective magnetic moment of the spherically
averaged $\Gamma _{7}^{(1)}$ CF ground state 
$\bar{\mu}_{eff}=\sqrt{2/3(\mu _{ab})^2+1/3(\mu _{c})^2}=1.621\mu _{B}$. 


\vskip 4cm 
\noindent
\begin{table}[tbp]
\epsfxsize=3.0in\epsffile{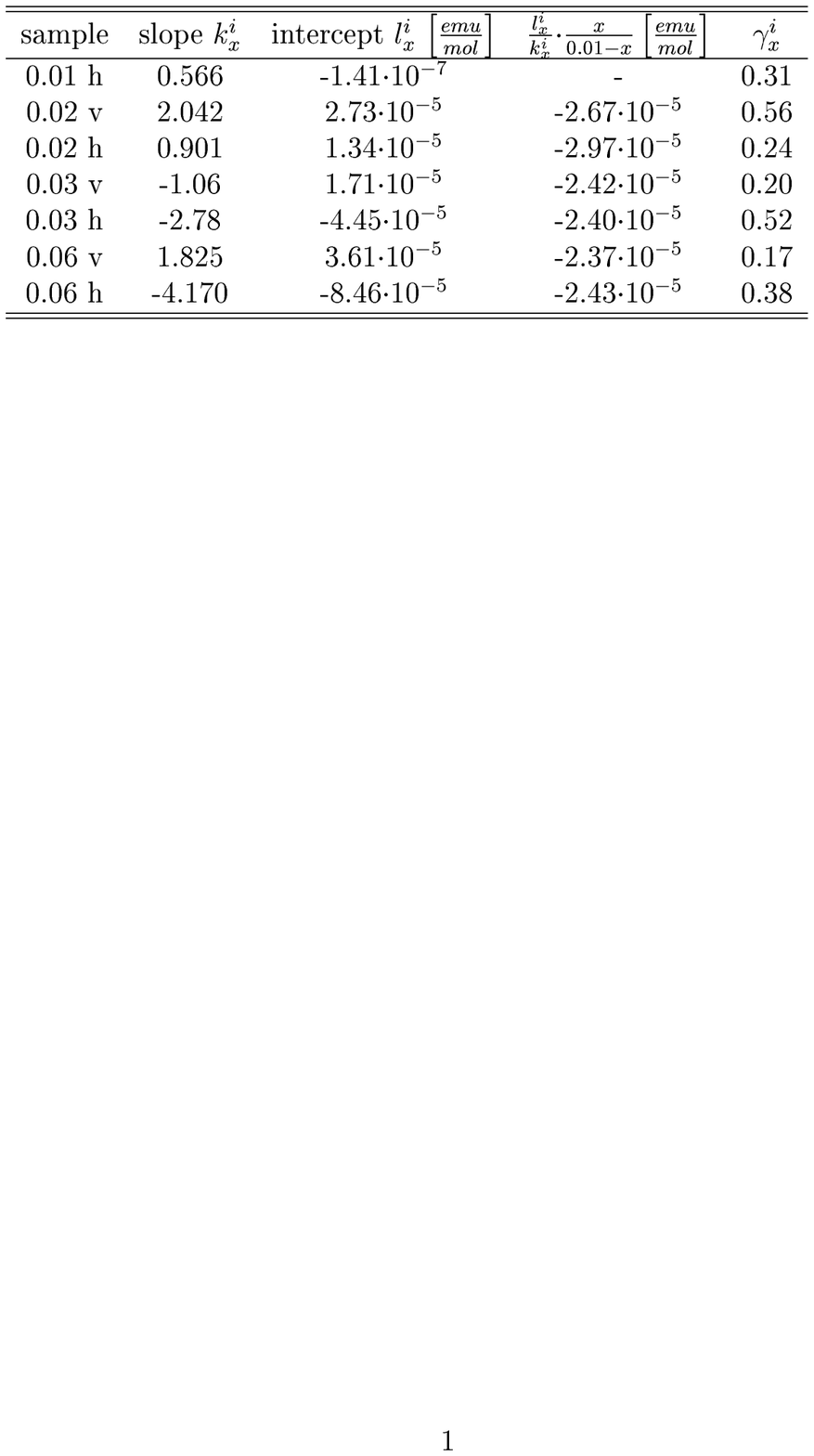}

\caption{The parameters from Fig.\ 9.}
                                          \label{tb1}
\end{table}


\begin{references}
\bibitem{systematic_error_1}
The samples with the same nominal concentration are obtained by 
powdering a small quantity of the material cut from different ends 
of the Ce$_{x}$La$_{1-x}$Cu$_{2.05}$Si$_{2}$ ingots.
\bibitem{systematic_error_2}
The inset in Fig.\ 1 shows the total susceptibility, rather than 
the impurity contribution. Thus, we can expect some deviation from 
the universal low-temperature behavior.
\bibitem{anizo_corr}
In Eq.\ (\ref{anizo-extract}) we replaced 
$\gamma_{1}\Delta \chi_B(0.01)$ by $\gamma_{1}\Delta \chi_0$,
i.e., we neglected the effects of the Ce doping on the background 
anisotropy of the sample with 1 at\% of Ce.
\end{references}
\end{document}